\documentclass{aa}  

\usepackage{graphicx}
\usepackage{txfonts}
\begin{document}

   \title{Analyzing temporal variations of AGN emission line profiles in the context of (dusty) cloud structure formation in the broad line region}

   \subtitle{}

   \author{J. Esser
          \inst{1,2}
          \and
          J.-U. Pott \inst{1}
          \and
          H.~Landt \inst{3}
          \and
          W.~D.~Vacca \inst{4}
          }

   \institute{Max Planck Institut f\"{u}r  Astronomie, K\"{o}nigstuhl 17, 69117 Heidelberg, Germany\\
              \email{esser@mpia.de}
              \and International Max Planck Research School for Astronomy \& Cosmic Physics at the University of Heidelberg, Germany\\
              \and Centre for Extragalactic Astronomy, Department of Physics, Durham University, South Road, Durham, DH1 3LE, UK\\
              \and SOFIA-USRA, NASA Ames Research Center, MS 232-12, Moffet Field, CA 94035, USA\\
             }

   \date{Received 20 September 2018; accepted 5 November 2018}

 
  \abstract
   {The formation processes and the exact appearance of the dust torus and broad line region (BLR) of active galactic nuclei (AGN) are under debate. Theoretical studies show a possible connection between the dust torus and BLR through a common origin in the accretion disk. However observationally the dust torus and BLR are typically studied separately. NGC~4151 is possibly one of the best suited Seyfert~1 galaxies for simultaneous examinations because of its high number of both photometric and spectroscopic observations in the past. Here we compare changes of the dust radius to shape variations of broad emission lines (BEL). While the radius of the dust torus decreased by almost a factor of two from 2004 to 2006 shape variations can be seen in the red wing of BELs of NGC~4151. These simultaneous changes are discussed in a dust and BEL formation scheme. We also use the BEL shape variations to assess possible cloud distributions, especially in azimuthal direction, which could be responsible for the observed variations. Our findings can best be explained in the framework of a dust inflated accretion disk. The changes in the BELs suggest that this dusty cloud formation does not happen continuously, and over the whole accretion disk, but on the contrary in spatially confined areas over rather short amount of times. We derive limits to the azimuthal extension of the observed localized BEL flux enhancement event.}

   \keywords{Galaxies: active - Galaxies: Seyfert -  galaxies: individual: NGC 4151 - quasars: emission lines
               }
   \titlerunning{Implications from shape variations of BELs on dust production in AGN and the structure of the BLR}
   \maketitle
%

\section{Introduction}

The modern standard model of active galactic nuclei (AGN) was established more than two decades ago
(\citealt{Antonucci1993,Urry1995}). While this model is still generally accepted, it has been refined in recent years.
While the dusty torus, which obstructs the view onto the central parts of the AGN depending on the inclination of the AGN, was assumed to be relatively homogeneous and symmetrical, there are findings suggesting a rather clumpy dust torus
consisting of a population of individual dusty gas clouds (e.g., \citealt{Nenkova2002,Hoenig2010}) similar to the broad line region (BLR) clouds. It is assumed that the toroidal distribution of dusty clouds has a
(somewhat extended) inner edge. The location of this inner edge is regulated, possibly by dust sublimation in the intense radiation field of the inner hot accretion disk. Gas clouds located in the BLR
rotate around the black hole at velocities up to a few 10000~$\rm{km~s^{-1}}$. Variations of both the overall flux and the
shape of the broad emission lines (BEL) have been reported (e.g., \citealt{Sulentic2000,Shapovalova2010,Ilic2015}). In all the publications mentioned above BELs at optical wavelengths were used, but here we have used infrared spectra. This is primarily due to data availability, but it also allows us to compare the shape variations found in the optical by \cite{Shapovalova2010} for NGC 4151 to shape variations in the infrared and prominent BELs such as Pa$\beta$ are not contaminated by emission lines of other chemical species (\citealt{Landt2011b}). The variations
of the shape of the broad emission lines can be explained by a changing and non-symmetrical distribution in azimuthal direction of those
clouds in the BLR or inflows and outflows of gas clouds. A theoretical description of the relation between cloud distribution in the BLR and the resulting BEL flux is presented in \cite{Stern2015}. However \cite{Stern2015} describe an azimuthally averaged temporal mean cloud distribution in radial direction, which we have modified for this article to allow for the localized temporal profile shape variations observed.

One of the promising applications of AGN is the use of the relation between the luminosity of the accretion disk and the radius of the dust torus or the inner edge of the BLR ($L~\propto~R^{0.5}$) as a standard ruler. In
numerous reverberation campaigns this relation was found to be true both for the dust torus (e.g.,
\citealt{Suganuma2006}) and the BLR (e.g., \citealt{Kaspi2000,Bentz2006}). Use of the relation at higher redshifts requires to establish the well-known size-luminosity relation for local lower luminosity Seyfert AGN to brighter quasars, which are observable at larger redshifts. 
\cite{Lira2018} determined the BEL lag of 17 quasars with a redshift up to $z~=~3$ and \cite{Hoenig2017} proposed to use AGN as standard candle for cosmology up to $z~=~1.2$ in the VEILS survey using dust time lags while the upper redshift limit for such studies is around four if BEL lags are used (e.g., \citealt{Watson2011}). 

While the size-luminosity relation works for samples, its scatter is significant, and in part due to the intrinsic astrophysical processes at work in the sub-parsec-scale AGN environment. \cite{Koshida2009} found
this relation for NGC~4151 as an overall trend but the evolution of the dust radius did not closely follow the evolution
of the luminosity of the accretion disk. This lead to a large scatter from the overall trend.

For the solution of this problem the production and destruction of dust in AGN in relation to accretion disk luminosity changes needs to be better understood. There are two competing classes of models for the dust production in AGN: Either the dust is produced by stars outside the AGN itself (e.g., \citealt{Schartmann2010}) or dust might be produced locally, as outflows from the accretion disk (\citealt{Czerny2011,Czerny2017}) or above the accretion disk directly at the location of the dust torus itself. The origin of the energy maintaining a geometrically thick toroidal dust structure is also debated. In contrast to radiation pressure dominated models, \cite{Bannikova2012} proposed a less dynamic model in which inclined orbits of the dust clouds and their self-gravity are responsible for the toroidal shape of the dust. In this model, the production of dust (to capture radiation pressure) is not necessary. However,  it is hard to explain the observed variations of the radius of the dust torus and shape variations of BELs without the production of additional clouds. This indicates that the
processes governing the radius of the dust torus as well as the formation mechanism of the broad line clouds deserves further investigation.

If the dust is not produced outside the AGN certain conditions have to be met regarding the temperatures produced by the radiation from the accretion disk: Dust sublimates at temperatures above at least 1500~K (e.g., \citealt{Barvainis1987,Schartmann2005,Nenkova2008}) while \cite{Baskin2018} suggest sublimation temperatures as high as 2000~K. On the other hand, dust can be created if the temperature drops below 
approximately 1000~K as shown for outflows of evolved stars where conditions similar to the BLR clouds are present (\citealt{Groenewegen2009}). Therefore the temperatures at the inner edge of the dust torus should be in this range and should be below 1000~K in order to produce dust.
\cite{Czerny2011} find that the inner edge of the BLR coincides with the point where the temperature inside the accretion disk drops
to around 1000~K while the temperature of gas clouds above the accretion disk is still higher due to the radiation originating from
the central parts of the accretion disk. From these results a model was constructed in which the dust is formed within the
accretion disk.Subsequently radiation pressure lifts the dusty gas clouds above the accretion disk where the dust in the clouds is sublimated by the accretion disk radiation and the clouds become visible as part of the BLR. Additionally the sublimation of dust leads to a lower opacity of the clouds and, due to gravity exerted by the accretion disk, they subsequently fall
back to the accretion disk. \cite{Baskin2018} refined this model by exploring the dust properties in the inner parts of the AGN. This way they described the distribution of dusty and dustless gas clouds within the \cite{Czerny2011,Czerny2017} model.

In this paper we describe a possible connection between the variability of the BELs shape of NGC~4151 to the changes of the
dust radius found by \cite{Koshida2009}. In Section~\ref{ch2} we describe how the BELs were extracted from the observed spectra
for our analysis. This is followed by the description of the BEL profile variability in Section~\ref{ch3} and
how we can model the observed BELs of NGC~4151 using (and expanding) the simple parametrization of the BLR 
from \cite{Stern2015} in Section~\ref{ch4}. In Section \ref{ch5} we discuss what these results suggest about how dust and the BLR is created in the inner AGN region. A summary of our findings will be given in Section~\ref{ch6}.

   \begin{figure*}
    \centering
    \includegraphics[width=17cm]{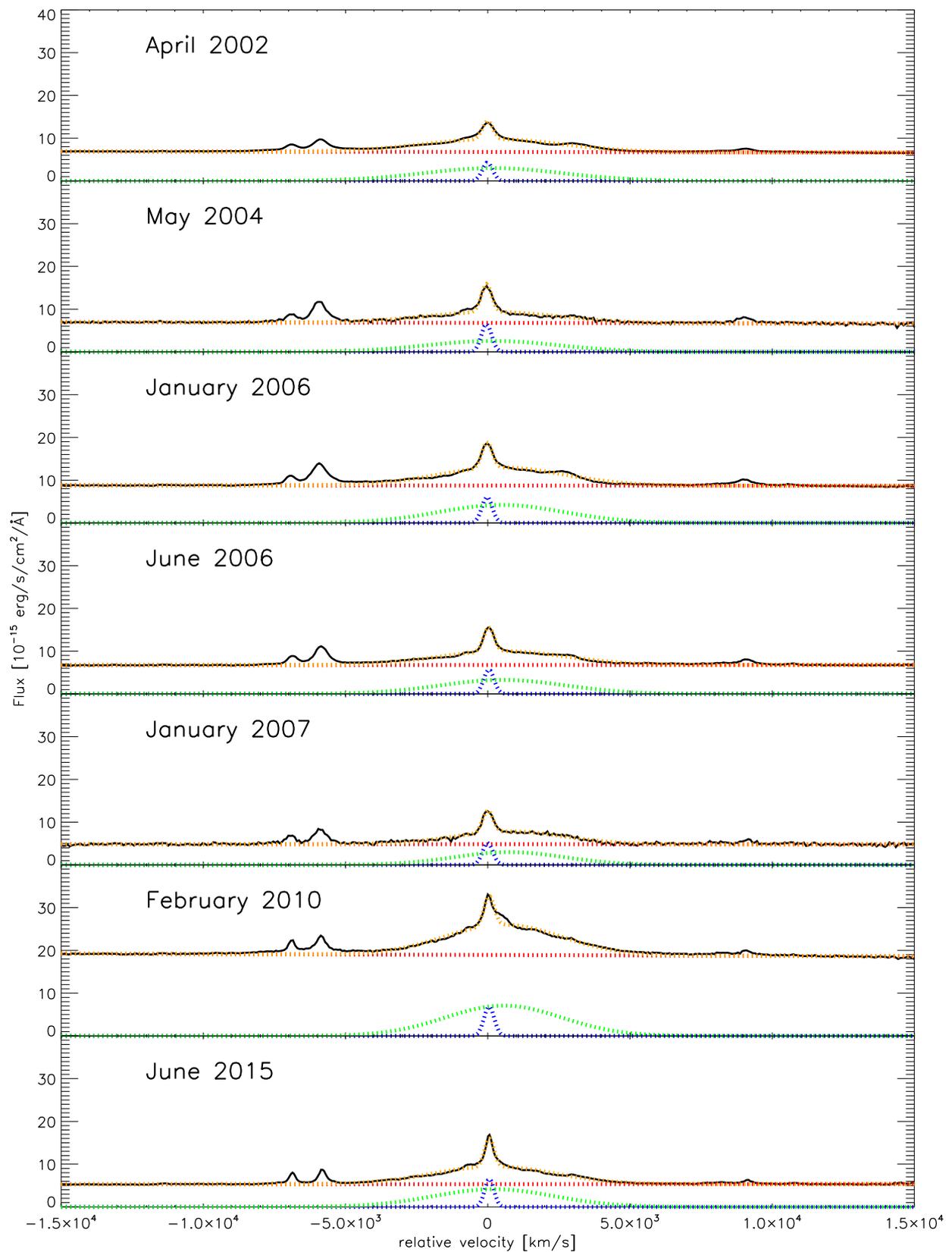}
    \caption{Environment around the Pa$\beta$ line in each spectrum. The data is shown in black along with the NEL fit (blue), BEL fit (green), continuum fit (red) and the sum of those fits (orange). }
    \label{figfit}
   \end{figure*}

   \begin{figure}
    \centering
    \includegraphics[width=\hsize]{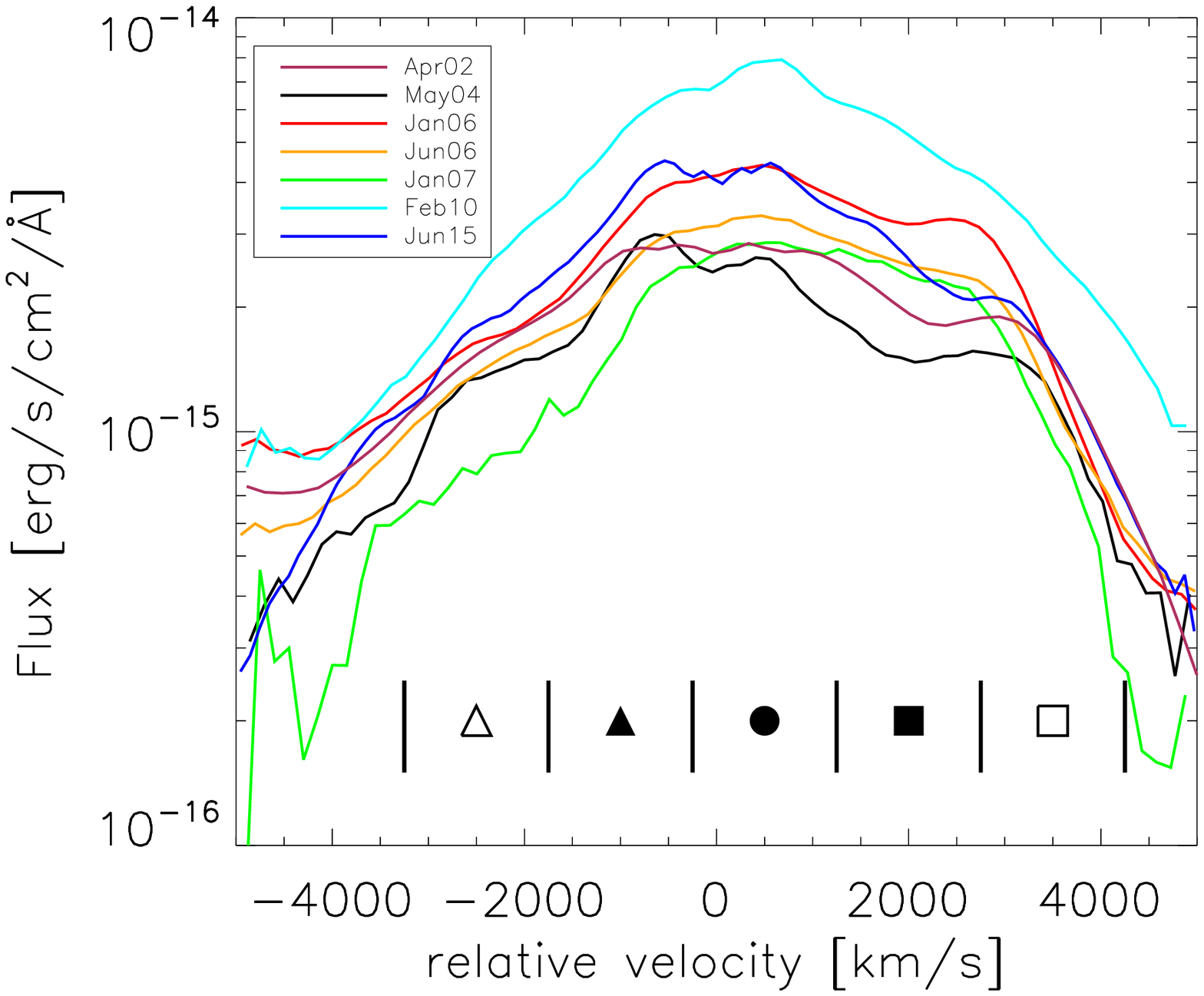}
    \caption{Pa$\beta$ BEL for our seven observation epochs. The color coding for the epochs is shown in the upper
    left corner of the plot. The subtraction of the NEL and continuum flux are described in Section~\ref{ch2}.
    Wiggles can be seen for most epochs approximately at the center of the BEL. We cannot rule out that these are residuals from the NEL
    subtraction. Therefore it is not possible to tell whether these shape variations originate in the BLR. At the bottom of the figure the velocity bins used for Fig.~\ref{figbinnedflux} are shown and
    color coded from blue to red. The strongest shape variations can be seen in the red shifted part of the BEL especially around velocities of 2500~$\rm{km~s^{-1}}$.}
    \label{figpabeta}
   \end{figure}


\section{Data reduction}
\label{ch2}

All spectra used here were taken with the SpeX spectrograph (\citealt{Rayner2003}) at the NASA Infrared
Telescope Facility (IRTF). They were obtained between April 2002 and June 2015 using the short cross-dispersed mode
(SXD, 0.8-2.4~$\rm{\mu}$m). All spectra were reduced using Spextool (\citealt{Cushing2004}). The 2002 spectrum was published by \cite{Riffel2006} and the next four spectra were used by \cite{Landt2008,Landt2011}. Information about the observations and the resulting spectra are given in Table~\ref{tableobs}. More details and
information on the reduction process are provided by \cite{Riffel2006} and \cite{Landt2008}. The sixth epoch from February 2010 has
already been used by \cite{Schnuelle2013}. The last spectrum was reduced in the same way as the sixth and
both spectra were calibrated as described by \cite{Vacca2003}. There is an additional spectrum from February 2015 from \cite{Wildy2016} but not much changes in it compared to the June 2015 spectrum and as the June 2015 spectrum has a much better signal to noise ratio (S/N, especially important for weaker BELs) we decided to not include this here.

For the selection of the lines used in this paper several criteria had to be met: First the BLR flux had to be
sufficiently high to assure S/N of at least 30 at the peak of the continuum subtracted BEL to enable us to see the shape variations of the BELs. Furthermore, in order to get a good fit of the continuum flux, a region
without any other emission lines had to be located close to the examined emission line. Emission lines with
wavelength above 2.4~$\rm{\mu}$m could not be taken into account as only the two newest spectra include
this wavelength range. In the end we were left with three emission lines best suited for our analysis which largely fulfilled these requirements. These three lines
were Pa$\beta$, \ion{O}{I}~844~nm and Br$\gamma$. However, the mentioned S/N could be reached only for the Pa$\beta$ line. For the other two lines the S/Ns are below 20. Therefore we largely focus on the Pa$\beta$ line in this paper but the other two lines are also mentioned, to point out the similar changes in those lines.

\begin{figure}
   \centering
   \includegraphics[width=\hsize]{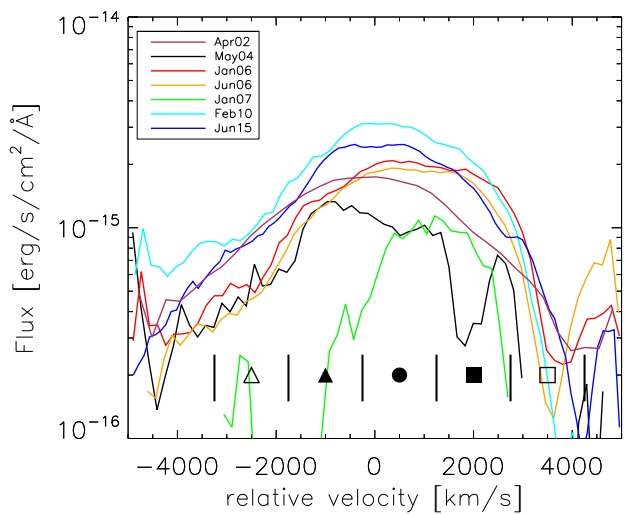}
      \caption{Same as Fig.~\ref{figpabeta} for the \ion{O}{I}~844~nm BEL.}
         \label{figoxygen}
   \end{figure}

\begin{figure}
   \centering
   \includegraphics[width=\hsize]{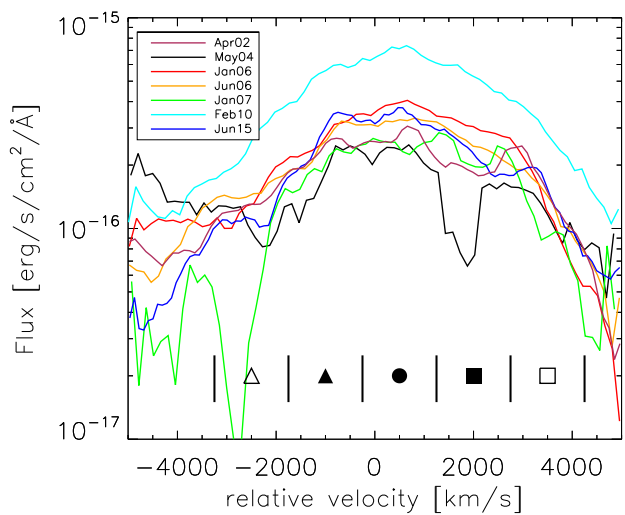}
      \caption{Same as Fig.~\ref{figpabeta} for the Br$\gamma$ BEL.}
         \label{figbrgamma}
   \end{figure}

In order to obtain the flux originating from the BLR the continuum flux as well as the flux from the narrow line region (NLR) had to
be subtracted. In order to get a good approximation of the continuum flux at the particular emission line we
apply a linear fit to a region which is free of emission lines from -14000 to -8000~$\rm{km~s^{-1}}$ and 8000 to 14000~$\rm{km~s^{-1}}$ from the emission line. In the case of Pa$\beta$ we can only
use a range between $\pm$10000 and $\pm$14000~$\rm{km~s^{-1}}$ due to \ion{Fe}{II} and \ion{S}{IX} emission around 1.26~$\rm{\mu}$m (visible in Fig.~\ref{figfit} to the left of the BEL). Flux from the NLR was
determined using a Gaussian fit. The results for the FWHM of that fit are shown in row 10 of Table~\ref{tableobs}. 
The fitted FWHM of the NEL profile is dominated by the effective resolution of the spectra and varies typically by only $\pm 1$ pixel around the mean. Only the April~2002 spectrum was an exception to this as the spectral resolution is a bit lower.  To minimize the influence of the BLR profile variability onto the NEL fit, we re-fit the spectrum with a NEL Gauss profile of constant width equal to the mean of the individual FWHM fits for all but the 2002 dataset.
In the case of the 2002 spectrum the individually measured FWHM was used. However the NEL flux was not constant for our spectra (most likely due to inaccuracies in the flux calibration) and was therefore left as a free parameter in the fits. The fits for each spectrum are shown for the Pa$\beta$ line in Fig.~\ref{figfit}. The resulting BEL plots with NEL and continuum subtracted are shown in Fig.~\ref{figpabeta} (Pa$\beta$),
Fig.~\ref{figoxygen} (\ion{O}{I}~844~nm) and Fig.~\ref{figbrgamma} (Br$\gamma$). In the velocity range of the NEL small bumps can be seen for each spectrum. In most cases it is hard to confirm whether these bumps are caused by shape variations of the emission lines. They could also be caused by residuals of the NEL fit. Especially in the February~2010 case a shape variation can be seen to the right of the NEL in Fig.~\ref{figfit}. So while it is possible that there are BEL shape variations around the NEL (i.e., at low velocities), we will concentrate on the variations in the red wings as they are unaffected by the NEL fit.

\begin{table*}
\caption{Journal of observations for our IRTF SpeX spectra.}             
\label{tableobs}      
\centering                          
\begin{tabular}{c c c c c c c c c c}        
\hline\hline                 
Date & MJD - 50000 & Exposure & Airmass & \multicolumn{3}{c}{Continuum S/N\tablefootmark{a}} & Mode & slit size & FWHM Pa$\beta$ NEL\tablefootmark{b} \\
     & & (s)      &         & J & H & K &  & & $\rm{km~s^{-1}}$ \\
\hline                        
   2002 Apr 23 & 2387 & 1800  & 1.10 & 189 & 303 & 406 & SXD & 0.8" x 15" & 700 \\
   2004 May 23 & 3148 & 720   & 1.13 & 32  & 45  & 121 & SXD & 0.8" x 15" & 560 \\      
   2006 Jan 08 & 3743 & 1920  & 1.07 & 103 & 212 & 345 & SXD & 0.8" x 15" & 490 \\
   2006 Jun 12 & 3898 & 1200  & 1.22 & 109 & 202 & 269 & SXD & 0.8" x 15" & 420 \\
   2007 Jan 24 & 4124 & 960   & 1.45 & 15  & 54  & 113 & SXD & 0.8" x 15" & 420 \\
   2010 Feb 27 & 5254 &       &      & 182 & 226 & 363 & SXD & 0.3" x 15" & 400 \\ 
   2015 Jun 23 & 7076 & 2520  & 1.09 & 150 & 163 & 250 & SXD & 0.3" x 15" & 400 \\
\hline                                   
\end{tabular}

\tablefoot{\tablefoottext{a}{S/N in the continuum over $\sim$~100~$\AA{}$ measured at the central wavelength of the J, H and K band.}
\tablefoottext{b}{This NEL FWHM was fitted with a Gaussian profile to the individual spectra. Part of the FWHM variation between the epochs is due to cross-talk between the NEL FWHM fit, and the variable BEL profile superposed onto the fixed NEL profile. The mean of the measured FWHM of the 2004 to 2015 spectra is 450~$\rm{km~s^{-1}}$.}}
\end{table*}

\section{BEL shape variation}
\label{ch3}

\subsection{Testing the temporal stability of the BEL profile}

The evolution of BEL profiles across the seven epochs are presented in Figs.~\ref{figpabeta} (Pa$\beta$), \ref{figoxygen}
(\ion{O}{I}~844~nm) and \ref{figbrgamma} (Br$\gamma$). Of these BELs, the Pa$\beta$ line has the highest flux and
S/N. The S/N of the \ion{O}{I}~844~nm line is further reduced
due to the higher continuum flux (so lower line-to-continuum-ratio) at these wavelengths compared to the Pa$\beta$ line and especially at the Br$\gamma$
line the continuum flux is much lower. Therefore, the further analysis is done for the Pa$\beta$ line alone while it is done for the sum
of the \ion{O}{I}~844~nm and Br$\gamma$ line to increase the S/N. The general findings are similar. 

Apart from the overall flux changes a variation of the BEL shape can be seen in the red wings. During the
first five epochs (from April 2002 to January 2007) a peak is apparent that flattens out with time and can no longer be
seen in the February 2010 epoch. In the June 2015 epoch a similar peak appears again at the red wing. The apparent variations around the center of the BELs were already briefly addressed at the end of Section~\ref{ch2}
and might be partly caused by residuals from the fit of the NELs. Our data does not sample strong shape variations in the blue wing. Similar emission peaks are found for the H$\alpha$ and H$\beta$ BELs of NGC~4151 by
\cite{Shapovalova2010} at the same time as our IR spectra were taken at the red wings and also at earlier times in the red wings only.

  \begin{figure}
   \centering
   \includegraphics[width=\hsize]{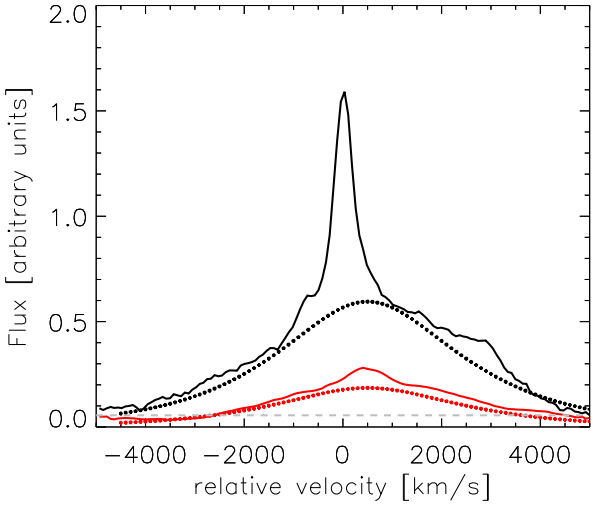}
      \caption{Mean (black solid line) and rms (red solid line) profile for the Pa$\beta$ line of all spectra despite the April 2002 spectrum (which had to be excluded here due to its lower spectral resolution). Along with this a model of a symmetrical BEL was plotted which we develop in Section~\ref{ch4} (red and black dotted lines) based on \cite{Stern2015}. The gray dashed line shows the median flux error of individual pixels at the Pa$\beta$ line.}
         \label{figrmsclassic}
   \end{figure}

In order to measure the significance of these variations we used the rms profile of our spectra. For the classical rms approach the fitted NEL flux was normalized in order to have a constant NEL flux for all spectra and the April 2002 spectrum had to be excluded due to its lower spectral resolution. The resulting mean and rms profiles are shown in Fig.~\ref{figrmsclassic} along with a symmetrical BEL model which we develop in Section~\ref{ch4}, based on the work of \cite{Stern2015}. In Fig.~\ref{figrmsclassic} it becomes apparent that the mean profile of our Pa$\beta$ emission line shows a significant shape discrepancy from the symmetrical profile in the red wing compared to the blue wing. Additionally the rms profile is slightly stronger in the red wing. If there were no shape variations it would be expected that the rms profile should be similar to the symmetrical profile. Nonetheless this approach of analyzing the rms spectral shape is not ideal to identify shape variations, since it mixes BEL flux variation with line profile shape variation.

As we look for shape variation alone in these profiles we need to normalize the spectra in order to get rid of the overall flux changes of the BELs. For this it is necessary to subtract the constant part of the emission line, namely the NEL, first. We then introduced a fit parameter (which acts as a normalization factor) for each spectrum ($p_i$) where the fit parameter for the May 2004 spectrum is fixed to 1.0. We then minimized the following rms profile:

\begin{equation}
 \label{eqrms}
 rms = \frac{\sum_{i=1}^6 (p_i F_i(\lambda) - \overline{F}(\lambda)) / (p_i \sigma_{F,i}(\lambda))^2}{\sum_{i=1}^6 (1/(p_i \sigma_{F,i}(\lambda)))^2}
\end{equation}

where $F_i(\lambda)$ is the BEL flux and $\sigma_{F,i}$ is the error of the BEL flux. $\overline{F}$ is the weighted mean flux of the seven spectra defined as

\begin{equation}
\label{eqrmsmean}
  \overline{F}(\lambda) = \frac{\sum_{i=1}^6 p_i F_i(\lambda)/(p_i \sigma_{F,i}(\lambda))^2}{\sum_{i=1}^6 (1/(p_i \sigma_{F,i}(\lambda)))^2}
\end{equation}

\begin{figure*}
\centering
\includegraphics[width=17cm]{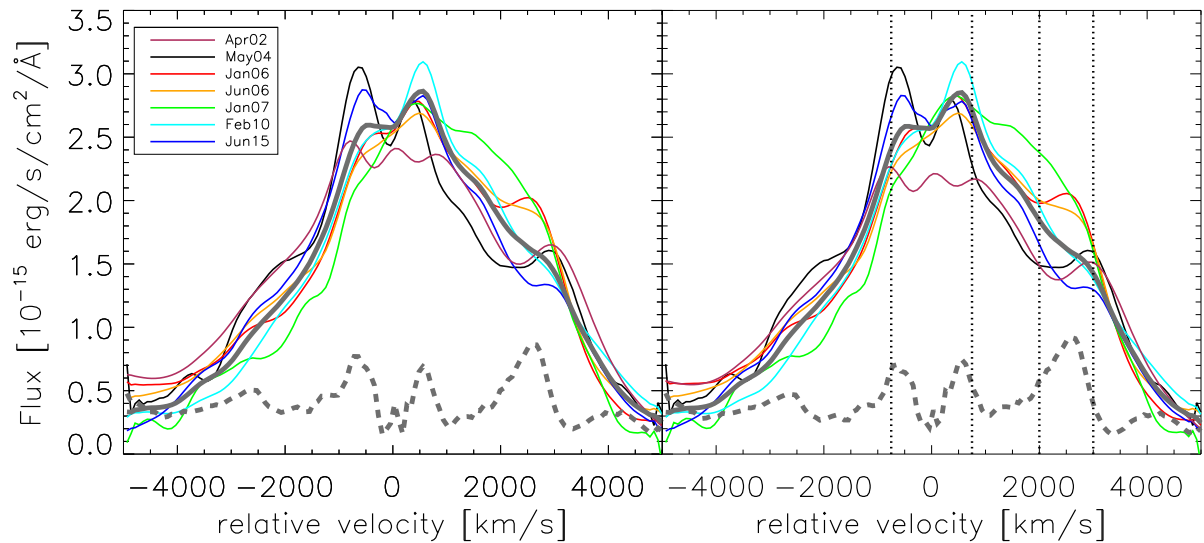}
\caption{Pa$\beta$ BELs normalized with the fit parameters from the rms minimization (Eq.~\ref{eqrms}) which are given in Table~\ref{tablerms}. The color coding of the BELs is given in the upper left corner. Additionally the mean spectrum (solid gray line) and the rms profile (dashed gray line) are shown. For better visibility the rms profile was multiplied by a factor of three. In the left plot the full BEL was used for the rms minimization. In order to make the minimization less sensitive to the shape variations of the BEL the indicated velocity ranges between -750 to 750~$\rm{km~s^{-1}}$ and 2000 and 3000~$\rm{km~s^{-1}}$ were excluded from the minimization in the right plot as the rms profile is largest in these regions indicating the strongest shape variations.}
\label{figrms}
\end{figure*}

At first we determine the rms profile using the complete BELs. The resulting fit parameters $p_i$ are shown in Table~\ref{tablerms} and the resulting rms (multiplied by a factor of three for better visibility) is shown in the left plot of Fig.~\ref{figrms} (dashed gray line) along with the spectra multiplied with their respective fit parameter (solid colored lines as in Fig.~\ref{figpabeta}) and the weighted mean Flux (solid gray line). Looking at this rms profile we find two velocity ranges now where the variations are strong. For the first one around 0~$\rm{km~s^{-1}}$ we can not easily tell to what extent these variations are real variations of the BELs or instead residuals from our NEL fit. Therefore we do not discuss them further here. The second peak of the rms profile is located in the region where the bumps discussed before can be seen from 2000 to 3000~$\rm{km~s^{-1}}$.

In the next step, we excluded those velocity ranges where the rms profile peak and repeat the process determining the rms profile as shown above. We do this as we want to include only the parts of the BELs which do not change their shape in the fitting process to avoid a biased flux normalization. The results of this second iteration are shown in the right plot of Fig.~\ref{figrms} and the fit parameters $p_i^*$ are given in Table~\ref{tablerms}. This does not strongly affect our fit parameters however $p_i$ is systematically smaller than $p_i^*$ with the exception of $p_0$ and with $p_4^*$ having the largest increase. Especially on the blue wing the spectra are pushed together by this change. However at velocities between 0 and 2000~$\rm{km~s^{-1}}$ the normalized flux of the May~2004 BEL (black line) is even further below the mean BEL while the normalized flux of the January~2007 BEL is even higher than the flux of the mean BEL. A similar behavior can be seen for the flux ratios compared to the May~2004 BEL. Excluding the varying parts of the BELs leads to a general increase in the flux ratios which is especially pronounced for the January~2007 BEL (compare rows 4 and 5 of Table~\ref{tablerms}).

We conclude that the BEL profile variation appears to be particularly pronounced between velocities of 2000 and 3000~$\rm{km~s^{-1}}$.

\begin{table}
\caption{Overview of the fit parameters obtained by the minimization of the rms profile (Eq.~\ref{eqrms}) and the flux ratios of the Pa$\beta$ BELs. In the second column the fit parameters are shown for the complete BEL and in the third column the fit parameters if the velocity range with high rms (compare Fig.~\ref{figrms}) are excluded. The flux ratios are shown for both of these cases as well with respect to the May~2004 BEL.}             
\label{tablerms}      
\centering                          
\begin{tabular}{c c c c c}        
\hline\hline                 
Date & $p_i$ & $p_i^*$ & $F_1/F_i$ & $F_1^*/F_i^*$ \\
\hline                        
   2002 Apr & 0.81 & 0.78 & 0.81 & 0.81 \\
   2004 May & 1.00 & 1.00 & 1.00 & 1.00 \\      
   2006 Jan & 0.60 & 0.63 & 0.62 & 0.65 \\
   2006 Jun & 0.76 & 0.79 & 0.78 & 0.81 \\
   2007 Jan & 0.89 & 0.95 & 0.98 & 1.08 \\
   2010 Feb & 0.38 & 0.38 & 0.39 & 0.40 \\
   2015 Jun & 0.61 & 0.63 & 0.63 & 0.66 \\
\hline                                   
\end{tabular}
\end{table}

\subsection{Time evolution of the BEL profile}

\begin{figure}
   \centering
   \includegraphics[width=\hsize]{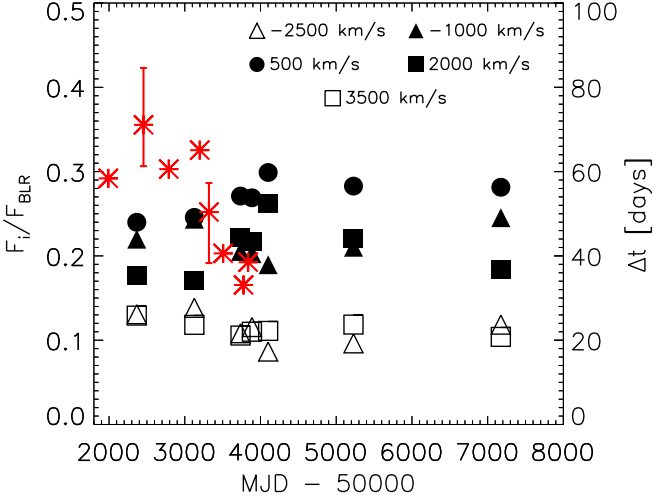}
      \caption{Fluxes in different velocity bins ($F_{ij}$) over the total flux from the Pa$\beta$ BEL ($F_{i}$) over time for the
      seven spectra. The different symbols indicate the different velocity bins which are also given
      at the bottom of Fig.~\ref{figpabeta} and their central velocity is given at the top of the plot. The error bars are smaller than the symbols except for two of dust radii where the error bars are given. Indicated with the
      red stars are the radii of the last five epochs determined by \cite{Koshida2009}. For those the
      reverberation delay in ld is plotted against time. At the same time the radius of the dust torus is reduced by a factor of approximately two, and the relative flux in the bin between 1250 and 2750~$\rm{km~s^{-1}}$ (orange crosses) increases significantly.
      }
         \label{figbinnedflux}
   \end{figure}

\begin{figure}
   \centering
   \includegraphics[width=\hsize]{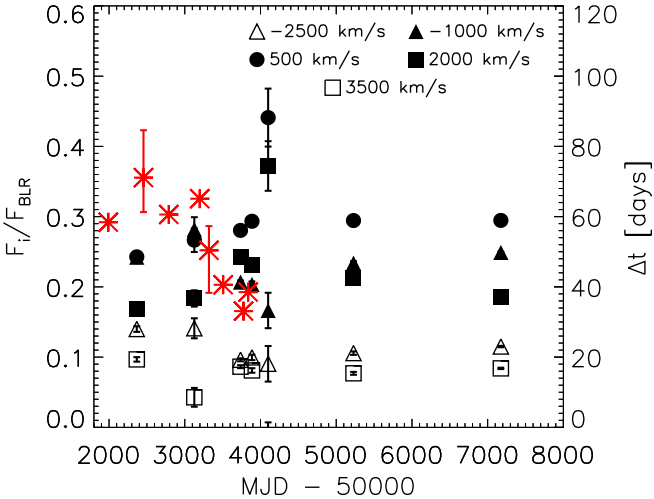}
      \caption{Same as Fig.~\ref{figbinnedflux} but for the sum of the ratios of the \ion{O}{I}~844~nm and the
      Br$\gamma$ BELs which are shown in Figs.~\ref{figoxygen} and \ref{figbrgamma}. We use the sum of the two BELs to increase the S/N as the strength of those lines is much weaker than the Pa$\beta$ line. A similar increase of the relative flux in the 1250 to 2750~$\rm{km~s^{-1}}$ bin (filled squares) can be seen for the \ion{O}{I}~844~nm and Br$\gamma$ BEL.}
         \label{figbinnedfluxcomb}
   \end{figure}

We divided our BELs in five equal velocity bins from -3250 to 4250~$\rm{km~s^{-1}}$ (indicated
by the symbols at the bottom of Figs.~\ref{figpabeta}, \ref{figoxygen}, and \ref{figbrgamma}). The bins are not centered around 0~$\rm{km~s^{-1}}$ as the BELs are shifted by approximately 500~$\rm{km~s^{-1}}$ with respect to the NELs. This velocity shift of the BELs with respect to the NEL center is also found by \cite{Shapovalova2010}. We then
integrated the flux in those bins as well as the overall flux from the BEL to get the ratio of those two values.
This ratio is plotted in Fig.~\ref{figbinnedflux} over time for the Pa$\beta$ BEL. Due to the lower S/N for the \ion{O}{I}~844~nm and Br$\gamma$ BELs we take the average of the ratios of the two lines for the same
plot (Fig.~\ref{figbinnedfluxcomb}). We decided not to take the sum of the fluxes to then get a combined average here as this would have lead to
a higher weighting of the \ion{O}{I}~844~nm line due to the higher flux of that BEL. The ratios are indicated by the
different symbols which are also given at the bottom of the figures showing the BELs (Fig.~\ref{figpabeta}, \ref{figoxygen}, and \ref{figbrgamma}).

It can be seen in Fig.~\ref{figbinnedflux} that the relative flux in the central bin and the bin centered around 2000~$\rm{km~s^{-1}}$ stays approximately constant between April 2002 and May 2004 and than gradually increases from the May 2004 BEL to the January 2007 BEL. The February 2010 BEL on the other hand shows an almost symmetrical BEL again (if 500~$\rm{km~s^{-1}}$ is assumed as the center of the BEL). This supports the result from the rms profile of strong shape variations between May 2004 and January 2007. Additionally it shows that this change does not evolve randomly but the BEL flux in the red wing steadily increases.

Another feature in the BELs appears similar to an absorption of flux at 2000~$\rm{km~s^{-1}}$ in the first period especially for the
\ion{O}{I}~844~nm and Br$\gamma$ BEL. For the Pa$\beta$ line it is not possible to tell whether the impression of a
slightly lower flux is caused by the enhanced flux at 3000~$\rm{km~s^{-1}}$. However as we definitely see this feature
in two out of three BELs at the same velocities, this could be a real absorption-like feature. It is also possible that this is caused by a lower emission at these velocities. However the rms is significantly higher above 2000~$\rm{km~s^{-1}}$ (compare Fig.~\ref{figrms}) and we therefore concentrate on the variations between 2000 and 3000~$\rm{km~s^{-1}}$. For the same reasons we neglect the slight shape variations in the blue wing. This shows that (slight) BEL shape variations are occurring throughout the BEL but are most significant between 2000 and 3000~$\rm{km~s^{-1}}$. There are some shape variations around 0~$\rm{km~s^{-1}}$ too, but these could be attributed to the NEL fit. Therefore, we cannot attribute these features cleanly to the BLR and we do not discuss them further in this paper.

\section{Modeling BELs}
\label{ch4}

Next we want to use the simple geometric parametrization of the BLR intensity distribution as described in \cite{Stern2015} to reproduce our BELs of NGC~4151. For the rest of 
the paper, we have used this approach to model the overall BEL profile, and 
extend it to realize the secondary peak observed. Of our seven spectra we chose the January 2006 spectrum as it is  the highest S/N spectrum taken in the time period showing line profile variability. Additionally it shows a significant bump in the region were the rms profile is the highest (Fig.~\ref{figrms}) and this spectrum was taken at a time when the radius of the dust torus determined by \cite{Koshida2009} was at a minimum. In total this BEL has the potential to provide the most information. The other BELs from this epoch show a similar overall width and a similar width of the bump (apart from the January 2007 BEL), although the bump intensity and location slightly changes. In order to do this we took Eq.~16 of \cite{Stern2015} describing the flux density of photons per unit velocity ($v$) that originate from the disk coordinate $(r, \varphi)$:

\begin{equation}
\label{eqStern1}
 \Phi _{v}^{*} (r, \varphi) = \frac{f(r)}{r} e^{-\frac{(v_{rot} \sin i)^2}{2 \sigma_v ^2}\bigg(\sin \varphi - \frac{v}{v_{rot}\sin i}\bigg)^2}.
\end{equation}

In this equation $f(r)$ describes the radial distribution of the line emission while the exponential function describes the local line broadening due to the non-rotational velocity components ($\sigma_v$) in the BLR with $i$ being the inclination between the accretion disk and the line of sight and $v_{rot}$ the rotational velocity (which is proportional to $r^{-0.5}$ as the gravitational field is dominated by the central black hole).

In Fig.~3 of \cite{Stern2015} the influence of $\sigma_v/(v_{rot} \sin i)$ on the shape of the BEL is shown. For values lower than  one, the BEL is double peaked and for values above one the shape becomes single peaked without significant further changes for larger $\sigma_v/(v_{rot} \sin i)$. Double peaked BELs are seen only if the FWHM reaches values above 10000~$\rm{km~s^{-1}}$ (e.g., \citealt{Eracleous2003}). Therefore we choose $\sigma_v/(v_{rot} \sin i)~=~1$. The radial distribution of line emission is described by $f(r)~\propto~r^1$ for $r~<~r_{BLR}$ and $f(r)~\propto~r^{-1}$ for $r~>~r_{BLR}$, where $r_{BLR}$ is the BLR radius measured with reverberation mapping. The changes of the BELs due to different radial distributions can be seen in Fig.~4 of \cite{Stern2015}. We choose this distribution as it is the widest of the given distributions in \cite{Stern2015} and reproduces the overall shape of our BELs well. Thus we do not overestimate the rotational velocity of the BLR clouds. Applying a steeper radial profile, for example $f(r)~\propto~r^{\pm 2}$), requires an increase of the rotational velocity by only 100~$\rm{km~s^{-1}}$.

The inner and outer radius of the luminous BLR clouds are adopted from \cite{Baskin2018} with values of $r_{in}~=~0.18~r_{BLR}$ and $r_{out}~=~1.6~r_{BLR}$. The choice of $r_{in}$ and $r_{out}$ slightly influences the shape of the BEL as well. For example, a larger $r_{out}$ leads to additional relatively slow clouds inducing a narrower BEL hence increasing the necessary rotational velocity to reproduce the width of our BELs. A larger $r_{in}$ has a similar effect because the fastest clouds are removed. 
The inner and outer radii were determined for a constant AGN luminosity by \cite{Baskin2018}. 
However, the optical lightcurve of NGC~4151 (e.g., \cite{Shapovalova2008,Koshida2009}) is not constant at all. Therefore BEL clouds might be present outside those radii depending on the earlier luminosities of NGC~4151 and the timescales on which BEL clouds are created and afterwards stop contributing to the BEL flux (for example by falling back to the accretion disk). As those timescales are not well understood, we use the radii from \cite{Baskin2018}.

\begin{figure*}
 \centering
 \includegraphics[width=17cm]{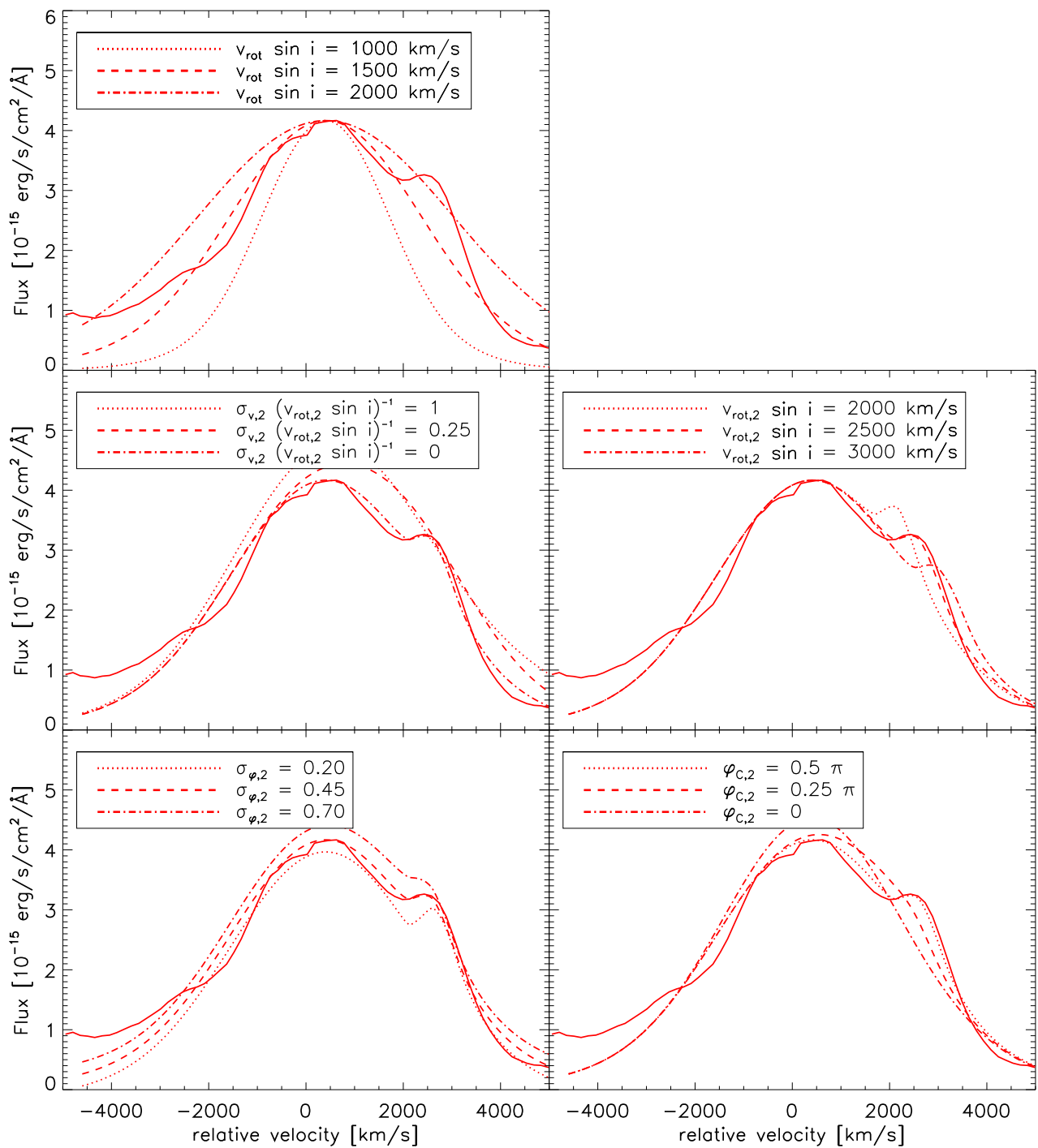}
 \caption{Influence of some of the parameter space of Eq.~\ref{eqStern3} on the shape of the BEL in comparison to the observed January 2006 Pa$\beta$ line (solid lines). In all plots the BEL was modeled with the parameters which reproduce the Pa$\beta$ line best. These are $v_{rot}~=~1500~\rm{km~s^{-1}}$, $\sigma_{v,2}~(v_{rot}~\sin~i)^{-1}~=~0$, $v_{rot,2}~=~2500~\rm{km~s^{-1}}$, $\sigma_{\varphi,2}^2~=~0.45~\rm{rad}$ and $\varphi_{C,2}~=~0.5~\pi$. The parameters varied are shown in the legends at the top of each plot. In the upper left plot the changes due to different rotational velocities are shown. The middle left plot illustrates how increasing the local line broadening leads to a wider bump. A visualization of the influence of a changing rotational velocity of the additional clouds is given in the right plot in the middle. In the lower left plot different widths of $g(\varphi)$ are shown. We note that the dotted line is shifted down and the dashed dotted line is shifted up by $0.2~\times~10^{-15}~\rm{erg/(s~cm^2~\mathring{A})}$ for better visibility. Finally in the lower left plot the center of the additional clouds ($\varphi_{C,2}$) is varied. These plots are not to be understood as fits to the data. Rather we want to explore how different parameters change the modeled BEL.}
 \label{figmodel1500}
\end{figure*}

This simple description of the BLR reproduces the overall shape of the Pa$\beta$ lines very well. As the overall shape and width of the Pa$\beta$ line does not change too much with time (compare Fig.~\ref{figrms}) we only show the comparison to the January 2006 Pa$\beta$ line in the upper left plot of Fig.~\ref{figmodel1500}. \cite{Shapovalova2010} also find that the FWHM of the BELs of NGC~4151 does not show strong changes (the FWHM only becomes smaller for a short time in 2000) despite the significantly reduced flux in a ten year span from 1996 to 2006. 

Next we want to reproduce the January 2006 spectrum with a separate peak around 2600~$\rm{km~s^{-1}}$ in addition to the main Gaussian profile. We note that the center of the Pa$\beta$ BEL is shifted by $v_{shift}~=$~500~$\rm{km~s^{-1}}$ with respect to the center of the narrow line. The modeled flux was normalized to match the maximum observed flux in all plots of Fig.~\ref{figmodel1500}.

This process should not be understood as a fit to the data, which is difficult to interpret since the parameters are partly correlated.  As we will show below, some of the elements of Eq.~\ref{eqStern1} have a similar effect on the BEL shape (e.g., the cloud distribution in $r$ and $v_{rot} \sin i$). Therefore a fitting process could not distinguish between those parameters any way. Rather we want to explore how the \cite{Stern2015} description of the BLR has to be tweaked in order to reproduce such a narrow bump, to explore the spatial information encoded in the BEL velocities.

Equation~\ref{eqStern1} can well reproduce the BEL  beside the bump and at velocities below -3000~$\rm{km~s^{-1}}$ with $v_{rot} \sin i~=~1500~\rm{km~s^{-1}}$ (upper left plot of Fig.~\ref{figmodel1500}). To be able to describe features in the BELs like the peak present at the red part of the January 2006 Pa$\beta$ line (compare also Fig.~\ref{figpabeta}) we need to vary the cloud distribution in azimuthal direction as well, adding a function $g(\varphi)$ to Eq.~\ref{eqStern1}:

\begin{equation}
\label{eqStern2}
 \Phi _{v}^{*} (r, \varphi) = \frac{f(r)}{r} (g(\varphi) + 1) e^{-\frac{(v_{rot} \sin i)^2}{2 \sigma_v ^2}\bigg(\sin \varphi - \frac{v}{v_{rot}\sin i}\bigg)^2}.
\end{equation}

We chose a Gaussian distribution centered around $\varphi_{C,2}~=~0.5~\pi$ with its width denoted by $\sigma_{\varphi,2}$ from hereon:

\begin{equation}
 g(\varphi) \propto e^{-\frac{(\varphi - 0.5 \pi)^2}{2 \sigma_{\varphi,2}^2}}.
\end{equation}

\begin{figure}
 \centering
   \includegraphics[width=\hsize]{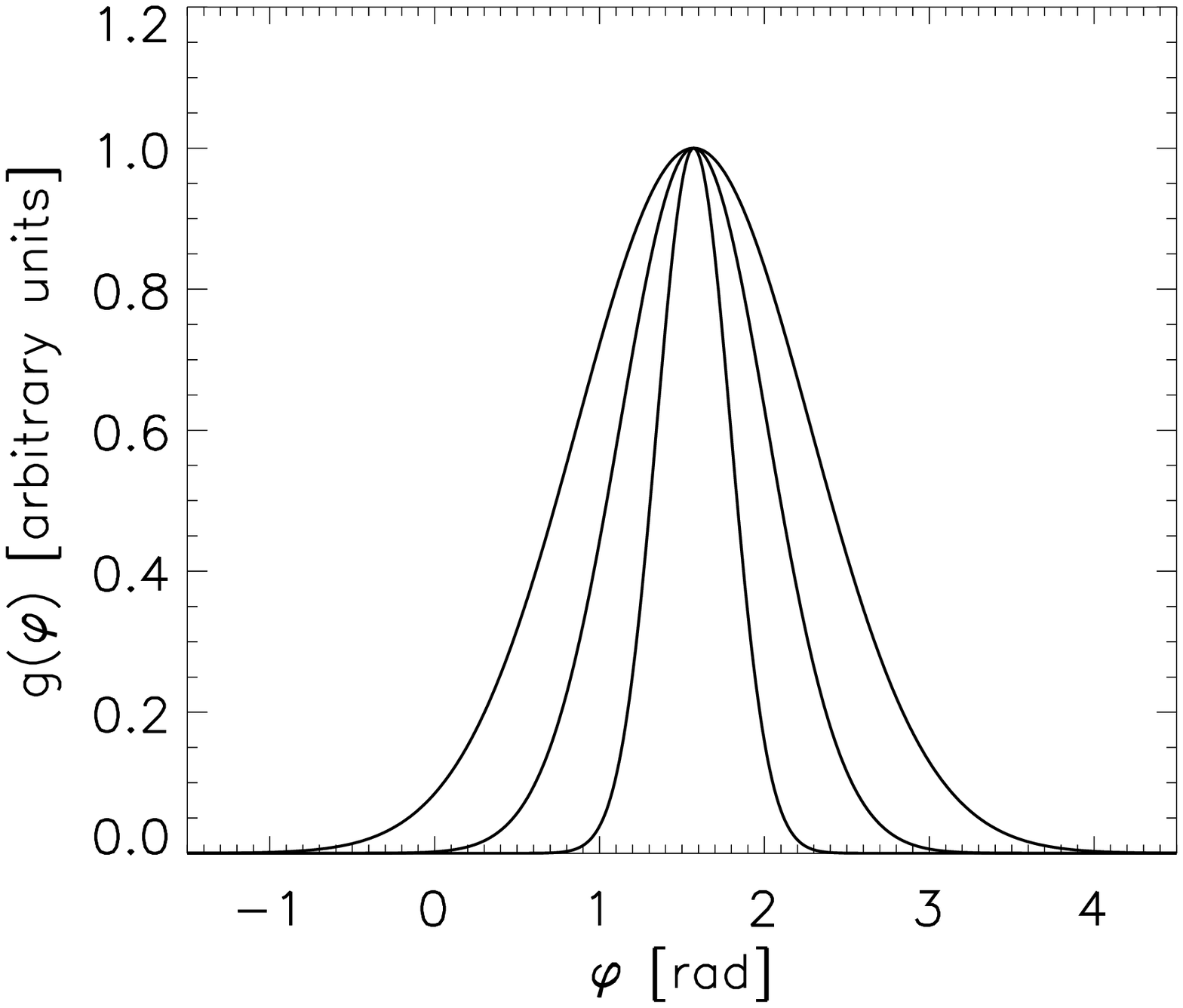}
   \caption{Distribution of additional clouds in the BLR introduced by $g(\varphi)$ in Eq.~\ref{eqStern2} and \ref{eqStern3} in azimuthal direction. These additional clouds are needed to reproduce the bump in the January 2006 spectrum in Fig.~\ref{figpabeta} (red line). They are modeled as a Gaussian distribution with a width of $\sigma_{\varphi,2}~=~0.20$, 0.45 and 0.70~$\rm{rad}$ and centered around $\varphi_{C,2}~=~0.5~\pi$.}
   \label{figazimuth}
\end{figure}

The distribution in $\varphi$ is shown in Fig.~\ref{figazimuth} for the different widths used in the lower left plot of Fig.~\ref{figmodel1500}. Apart from the lower left plot we always choose the width of the additional bump to be $\sigma_{\varphi,2}^2~=~0.2~\rm{rad^2}$ in Fig.~\ref{figmodel1500}.

However the local line broadening acts as a convolution on $g(\varphi)$. Therefore however small we choose the region in $\varphi$, where gas clouds in addition to the symmetric distribution are located, the contribution to the overall flux extends over a larger range of velocities compared to the January 2006 peak. This effect is shown in the middle left plot of Fig.~\ref{figmodel1500}. For values of $\sigma_{v,2}/(v_{rot,2}~\sin~i)$=~0.25 and 1 (dotted line and dashed line) the additional flux extends over a large velocity range. Only if $\sigma_{v,2}/(v_{rot,2}~\sin~i)$ goes to zero we can reproduce the bump properly. Hence we have to change the local line broadening term for the clouds described in $g(\varphi)$:

\begin{equation}
\label{eqStern3}
\begin{split}
 \Phi _{v}^{*} (r, \varphi) = \frac{f(r)}{r} e^{-\frac{(v_{rot} \sin i)^2}{2 \sigma_v ^2}\bigg(\sin \varphi - \frac{v}{v_{rot}\sin i}\bigg)^2} \\
 + \frac{f_2(r)}{r} g(\varphi) \delta \bigg(\sin \varphi - \frac{v}{v_{rot,2}\sin i}\bigg).
\end{split}
\end{equation}

A summary of the parameter space of this equation is given in Table~\ref{tableparams}.

\begin{table*}
\caption{Summary of the parameter space of our BEL modeling described in Eq.~\ref{eqStern3}. The parameters are given together with the values we used to reproduce Fig.~\ref{figmodel1500} and a short description. Some of these parameters were varied in Fig.~\ref{figmodel1500}.}             
\label{tableparams}      
\centering                          
\begin{tabular}{c c c}        
\hline\hline                 
\multicolumn{3}{c}{BEL cloud distribution} \\
Label & Best value & Description \\                        
\hline                                   
$f(r)$               & $r^{\pm 1}$    & radial distribution of BLR emission \\
$f_2(r)$             & $r^{\pm 2}$    & radial distribution of BLR emission for additional clouds \\
$r_{in}$             & $0.18~r_{BLR}$ & inner radius of BLR (\citealt{Baskin2018}) \\
$r_{out}$            & $1.6~r_{BLR}$  & outer radius of BLR (\citealt{Baskin2018}) \\
$g(\varphi)$         & Gauss function & azimuthal distribution of BLR emission for additional clouds \\
$\varphi_{C,2}$      & $0.5~\pi$      & center of the Gauss function \\
$\sigma_{\varphi,2}$ & 0.45           & width of the Gauss function \\
\hline
\multicolumn{3}{c}{BEL cloud velocities} \\
Label                               & Best value     & Description \\
\hline
$v_{rot}$                           & $1500~\rm{km~s^{-1}}$ & rotational velocity of BLR clouds \\
$\sigma_v/(v_{rot} \sin i)$         & 1                     & velocity dispersion or local line broadening term of BLR clouds \\
$v_{rot,2}$                         & $2500~\rm{km~s^{-1}}$ & rotational velocity of additional BLR clouds \\
$\sigma_{v,2}/(v_{rot,2} \sin i)$   & 0                     & velocity dispersion or local line broadening term of additional BLR clouds \\
\hline
\end{tabular}
\end{table*}

\begin{figure}
 \centering
   \includegraphics[width=\hsize]{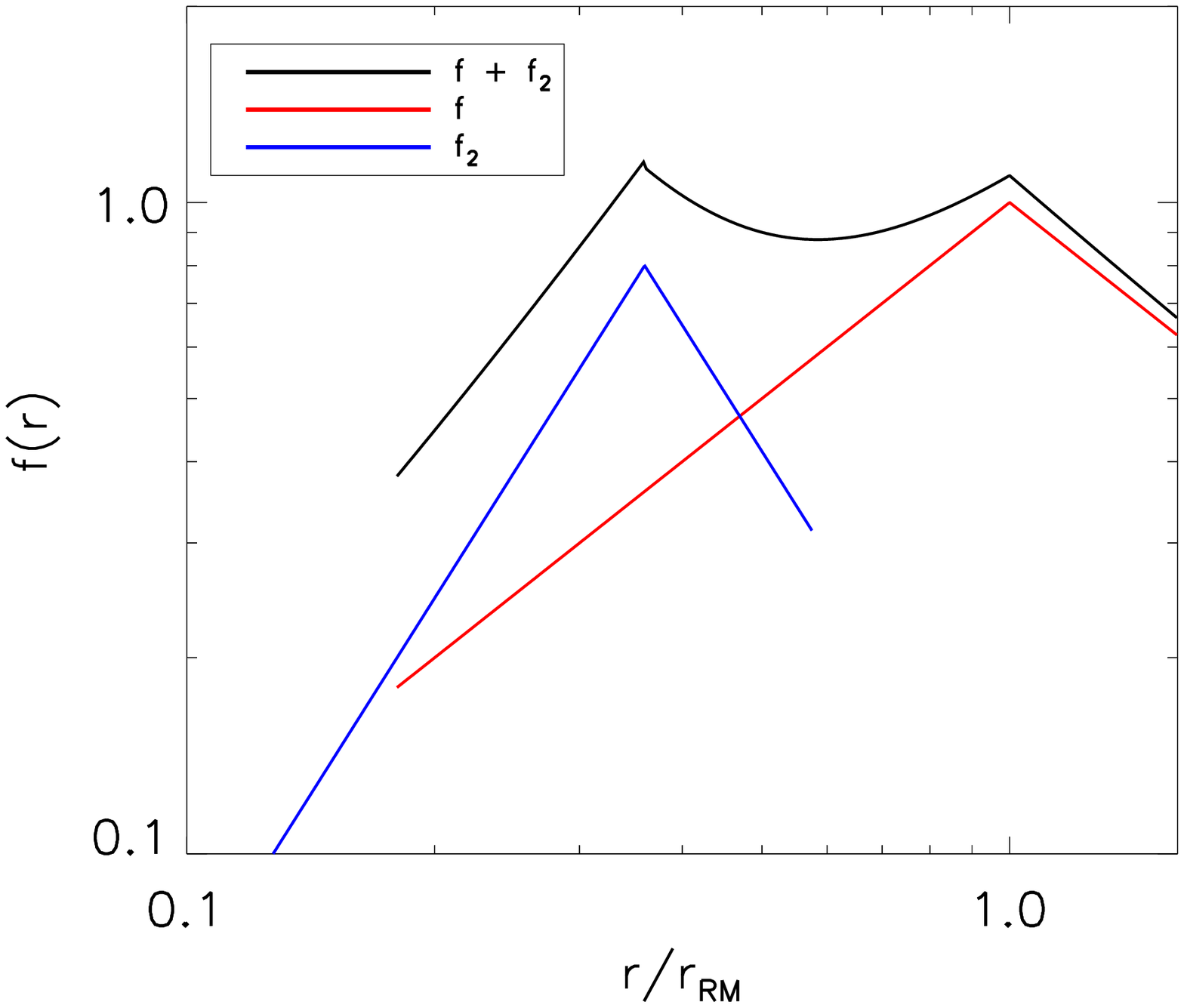}
   \caption{Radial distribution of line emission as used in Eq.~\ref{eqStern3} for $f$ (red line), $f_2$ (blue line) and the sum of the two components at $\varphi~=~0.5~\pi$ (where we put the additional clouds in azimuthal direction). $r_{rm}$ is the center of the overall BLR and the additional clouds are shifted inwards according to the ratio of $v_{rot}$ and $v_{rot,2}$ squared. Together with the distribution in azimuthal direction (compare Fig.~\ref{figazimuth}) the additional clouds are responsible for approximately 5~\% of the total BEL flux.}
   \label{figradial}
\end{figure}

The local line broadening does not have to vanish altogether. In fact the width of the peak depends as well on the distribution of clouds in radial and azimuthal direction. But with the delta function we get an upper limit on the volume of the additional clouds in $\varphi$. With the same argument, we changed the distribution in r to a steeper distribution with $f_2(r)~\propto~r^2$ for $r~<~r_{BLR}$ and $f_2(r)~\propto~r^{-2}$ for $r~>~r_{BLR}$. The radial distribution of line emission is shown in Fig.~\ref{figradial} for $f$, $f_2$ and the sum of the two components at $\varphi_{C,2}$. The flux originating from this additional bump is responsible for about 5~\% of the total BEL flux.

Even if we chose these clouds to be located at $\varphi_{C,2}~=~0.5~\pi$ where $v_{rot}~\sin~i$ is directed exactly away from us (the unprojected velocity is proportional to $\sin \varphi$) we would not be able to reach velocities above 1500~$\rm{km~s^{-1}}$. However the peak can be best reproduced with a velocity of 2500~$\rm{km~s^{-1}}$ as shown in middle right plot of Fig.~\ref{figmodel1500}. Therefore the additional clouds responsible for the bump emission need to be located closer to the black hole than $r_{BLR}$ as a rotational velocity of at least $v_{rot}~\sin~i~=~2500$~$\rm{km~s^{-1}}$ is needed to create a peak at this velocity. Therefore, a second velocity must be introduced in Eq.~\ref{eqStern3} ($v_{rot,2}$) which is the velocity of the additional clouds added with $g(\varphi)$.

In the lower left plot of Fig.~\ref{figmodel1500}, we show how changes of the width of $g(\varphi)$ changes the appearance of our bump. Apart from the peak becoming slightly wider with increasing width it does not change too strongly. The reason for this is the non-existing local line broadening. As long as the local line broadening is turned off, a constant cloud distribution in azimuthal direction leads to a double peaked BEL. This means the distinct peak around $v_{rot,2}~\sin~i$ will not show major changes to its width however big we choose $\sigma_{\varphi,2}^2$. Yet $\sigma_{\varphi,2}~\approx~0.45~\rm{rad}$ gives us the best resulting peak and if the additional clouds would be extended throughout the accretion disk there should be a second peak at the blue shifted wing of the BEL.

The results for a varying $\varphi_{C,2}$ is shown in the lower right plot. Moving the center towards 0 has two effects: The resulting peak widens and the center of the peak moves towards lower velocities. But unless $\sigma_{\varphi,2}$ is not very small the velocity change of the peak is smaller than $\sin~\varphi_{C,2}$ would suggest due to the convolution with the local line broadening term. However if we choose $\sigma_{\varphi,2}$ too small it can be hard to explain how to produce 5~\% of the BEL flux in such a small space on timescales of at most two years. For $\varphi_{C,2}~<~0$ the same peaks appear on the blue wing of the BEL.

There are three effects which could lead to the subsequent flattening of the peak in June 2006 and January 2007: The clouds rotating away from $\varphi_{C,2}~=~0.5~\pi$ (which would take longer than one year at these rotational velocities), turning on the local line broadening again (the clouds of $g(\varphi)$ no longer move with a common velocity) and additional similar events at lower projected velocities or a combination of these two effects. A further extend of the clouds in azimuthal direction can not explain this broadening alone as the small local line broadening always leads to a peak around $v_{rot,2}~\sin~i$ (compare the lower right plot of Fig.~\ref{figmodel1500}). The enhanced flux in January 2007 extends to around the center of the BEL. As we can see in the upper right plot of Fig.~\ref{figmodel1500} this can be realized with $\sigma_{\varphi,2}^2~=~0.2~\rm{rad^2}$ and local line broadening term of $\sigma_{v,2}~(v_{rot}~\sin~i)^{-1}~=~0.25$.

\section{Discussion}
\label{ch5}

In this chapter, we explore if the so far purely phenomenological description of BEL profile variability in NGC~4151 matches our understanding of physical processes occurring in that very central region around the SMBH. 

\subsection{Dust production}

In Fig.~\ref{figbinnedflux}, we show the dust radii of NGC~4151 (red stars) determined by \cite{Koshida2009} along with the increasing relative flux in the red wing of the BELs. This parallel change of BEL shape and dust radius in NGC~4151 raises the question whether this is coincidental or
is caused by a connection between the BLR and the dust torus and what this can tell us about dust occurrence and creation in AGN. An important property for the dust creation is the temperature of the dust torus. Only below a certain temperature dust can be created. This limiting temperature is usually assumed to be around 1000~K (e.g., \citealt{Czerny2011}) as this temperature was found for dust production in outflows of evolved stars where similar conditions are present as in BLR clouds (\citealt{Groenewegen2009}). Unfortunately \cite{Koshida2009} could not measure the temperature directly as they only had single band infrared fluxes
in the K band. However applying equation~1 from \cite{Kishimoto2007} the dust temperature can be connected to
the dust radius ($R_{sub,theo}$), the UV luminosity ($L_{UV}$), the sublimation temperature ($T_{sub}$) and the grain size of the dust ($a$):

\begin{equation}
 R_{sub,theo} = 1.3 \bigg(\frac{L_{UV}}{10^{46}\rm{erg/s}}\bigg)^{1/2} 
 \bigg(\frac{T_{sub}}{1500\rm{K}}\bigg)^{-2.8} \bigg(\frac{a}{0.05\rm{\mu m}}\bigg)^{-1/2}~\rm{pc}.
\end{equation}

Assuming a constant dust grain size and sublimation temperature for NGC~4151 we can calculate mean dust temperatures at different radii if we
assume certain temperatures in one of the periods from \cite{Koshida2009}. Our chosen period is the third one, as its luminosity is the highest while the dust radius is comparable to the first period where the luminosity is much lower. Therefore we can assume that the dust temperature should not be very low in the third period. We want to see
which temperature a dust cloud at the smallest dust radius found by \cite{Koshida2009} (33~ld) would have during
all periods:

\begin{equation}
 \label{eqTfig}
 T_{\rm{33 ld}} = T_{3} \bigg(\frac{R_{sub,theo,3}}{33~\rm{ld}}\bigg)^{1/2.8} 
 \bigg(\frac{L_i}{L_3}\bigg)^{1/5.6}.
\end{equation}

   \begin{figure}
   \centering
   \includegraphics[width=\hsize]{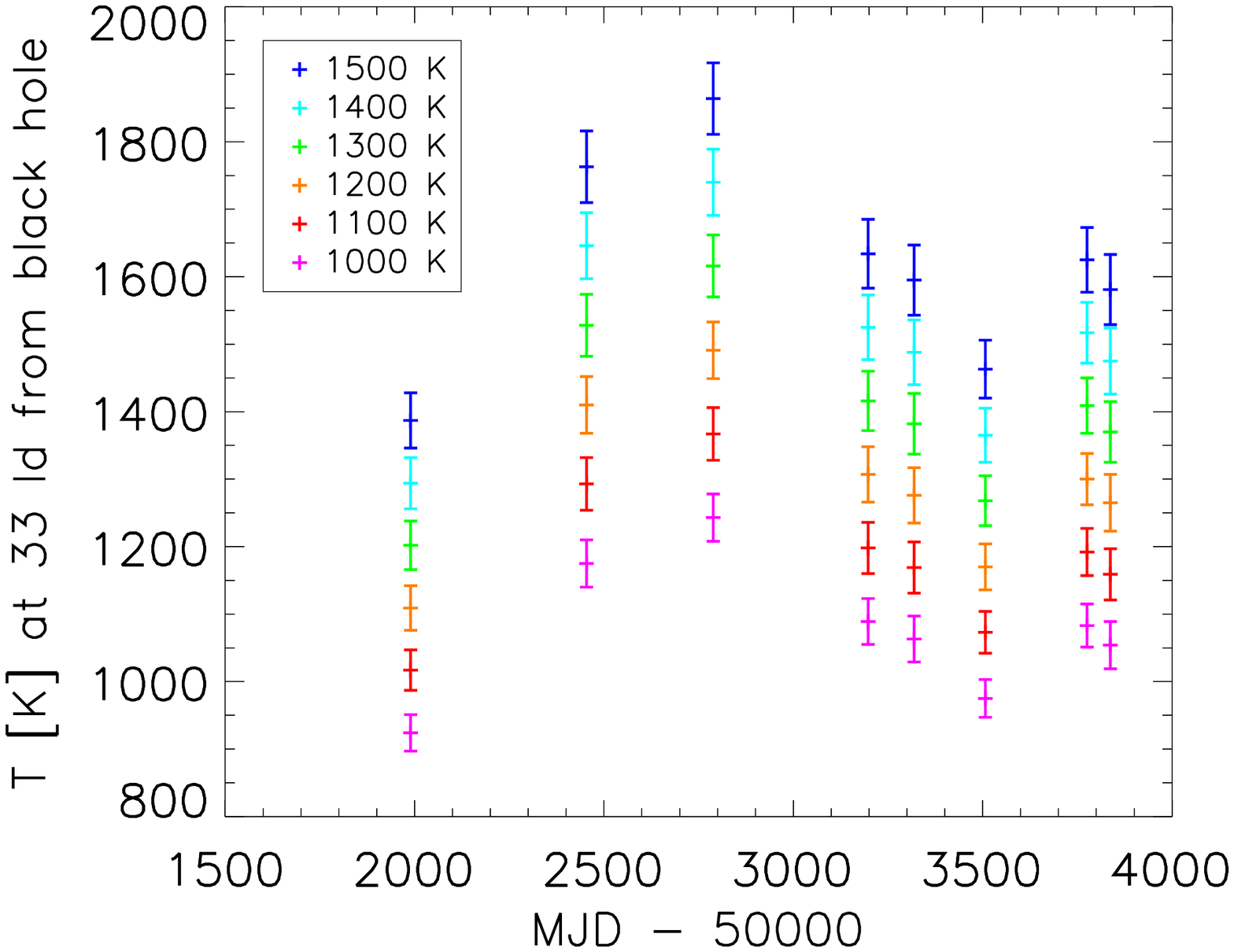}
      \caption{Different dust temperatures for NGC~4151 at a distance of 33~ld from the black hole assuming mean
      temperatures between 1000 and 1500~K during the third epoch of the NGC~4151 dust reverberation campaign
      from \cite{Koshida2009}. The temperatures were calculated using Eq.~\ref{eqTfig} with the assumption of a
      constant dust grain size. The color coding for the assumed mean temperatures during the third reverberation
      period is shown in the upper left corner of the plot. This shows that it is very hard to reach temperatures below 1000~K at a distance of 33~ld from the black hole, which are needed for dust production above the accretion disk.}
         \label{figtemp}
   \end{figure}

In this equation $T_3$ is the assumed mean temperature in the third period, $R_{sub,theo,3}$ is the sublimation
radius during the third period and $L_i$ is the mean luminosity in different epochs. With Eq.~\ref{eqTfig} we get the temperature at different times at a distance of 33~ld to the black hole.

The result of this line of thought is shown in Fig.~\ref{figtemp} for $T_3$ between 1000 and 1500~K. During the second period \cite{Landt2015} measured a dust temperature of 1316~K. During the third epoch the luminosity is higher while the dust radius stays almost the same. Therefore dust temperatures should be higher than 1300~K. Additionally the temperatures measured by \cite{Schnuelle2015} and \cite{Landt2015} for NGC~4151 never drop below 1200 K. \cite{Baskin2018} show that depending on especially gas density, grain size, and dust type $T_{sub}$ can reach temperatures as high as 2000~K. Therefore the temperature $T_3$ can be
assumed to be at least higher than 1300~K.

This means it is very hard to create dust above the accretion disk at the radii where dust reverberation finds it. In Fig.~\ref{figtemp} it can be seen that a
temperature below 1000~K can only be reached if $T_3$ is as low as $\sim$~1000~K and only in the sixth period
(apart from the first period). Otherwise the temperatures are well above this threshold for dust production at a
distance of 33~ld from the black hole. If $T_3$ is indeed higher than 1300~K the mean temperatures at this distance to the black hole can
not be much smaller than 1200~K. Therefore it is only possible to form the dust above the accretion disk if dust
formation can happen on very short timescales when the temperature is (significantly) below the mean temperature. If the dust were to be produced by stars outside the AGN as proposed by \cite{Schartmann2010} in NGC~1068 the decrease of the dust radius of $\sim$~30~ld in $\sim$~600~days would suggest a velocity of the dusty gas clouds towards the black hole of approximately 5~\% of the speed of light. This seems to be too high for an inflow as only ultra fast outflows from AGN have been observed in this velocity range. Velocities in a range up to 30~\% the speed of light peaking around 0.1~\% the speed of light were reported for these outflows (e.g., \citealt{Chartas2003,Tombesi2010}).

A solution to this problem (finding dust where it is too hot to form it) is provided by the model of the outer BLR formation from \cite{Czerny2011}. The dust is formed inside
the accretion disk where the radiation from the inner accretion disk is blocked by the accretion disk and
therefore temperatures can be lower than above the accretion disk where the clouds are directly exposed to the accretion disk. The dusty clouds are than pushed above the
accretion disk where the clouds are directly exposed to the radiation from the inner accretion disk and subsequently
destroyed by the radiation if the clouds are heated above the sublimation temperature. If the dust is indeed destroyed the gas clouds can
become visible as part of the BEL.

This relates to the changes of the dust radius and the shape of the BEL as follows. Dust clouds are produced within a range of radii in the the 
accretion disk smaller than the previously smallest dust radius (sublimation radius) above the accretion disk due to the additional radiation shielding inside the disk. Of these emerged dusty clouds the innermost clouds will loose their dust due to sublimation quickly and appear as addition to the BLR, while the dusty clouds will survive longer at small radii. If the amount of dust clouds is high enough, or the accretion disk luminosity drops at the same time these dusty clouds
will lead to a reduced time lag of the dust reverberation radius as seen by \cite{Koshida2009}. Thus the same cloud formation process in a radiation driven wind can add both to the BLR gas clouds, and the dusty clouds contributing to the torus.

\subsection{Kinematic features and possible origin of additional clouds}

The results from our BEL modeling (compare with Section~\ref{ch4}) support the view discussed above. The radius at which the shape of the BEL changes has to be at least two times smaller than $r_{BLR}$ of the overall BEL because of the velocities of the overall BEL and the additional peak. This is comparable to the changes in dust radius found by \cite{Koshida2009}. The enhanced flux can only be seen around 2100~$\rm{km~s^{-1}}$ in the BELs while the overall shape of the BELs does not change much with rotational velocities of around $v_{rot}~\sin~i~=~1500$~$\rm{km~s^{-1}}$. An explanation for this is that the BEL changes only in a confined area in azimuthal direction of the accretion disk and the overall distribution changes take much longer. It could be possible to explain the enhanced BEL gas density by an inflow of gas clouds, but in this case the changes in the shape on timescales of half a year from January 2006 to June 2006 and January 2007 are hard to explain.

The overall distribution of BEL clouds would not be symmetrical under these assumptions. In Fig.~\ref{figmodel1500} we showed that a broader distribution of additional clouds still leads to a distinct peak as long as the local line broadening term is negligible. Therefore it is hard to pinpoint how broad the distribution of additional clouds is as the width in flux of the additional peak is not very sensitive to this distribution. Nonetheless a width of $\sigma_{\varphi,2}~\approx~0.45~\rm{rad}$ can best reproduce the peak found in the January 2006 BEL and if we increase the local line broadening term again we can reproduce a wider additional flux without changing $\sigma_{\varphi,2}$ as seen in the January 2007 Pa$\beta$ BEL. In 2010 the bump vanished completely which can be caused by a combination of two effects: a further increase of the local line broadening term and a dispersion of clouds in azimuthal direction due to interactions with other BLR clouds. While there is no chance to get this azimuthal information of the distribution of dust clouds \cite{Schnuelle2015} found indications of a radial extend of the inner edge of the dust torus as well in NGC~4151 (although not at the same time). It could be interesting to see whether simulations can reproduce the changes in dust radius even if the 'new' dust has only a similarly small extend in azimuthal direction.

We can also infer an absolute value of $r_{BLR}$ for the two velocities we got from our modeling using the black hole mass determined by \cite{Grier2013} of $M_{BH}~=~3.62^{+0.39}_{-0.22}~\times~10^7~\rm{M_{\odot}}$. This leads to $r_{BLR,2}~=~30^{+3}_{-2}~\rm{ld}~\sin^2~i$ for $v_{rot,2}~\sin~i~=~2500~\rm{km~s^{-1}}$ and $r_{BLR}~=~83^{+9}_{-6}~\rm{ld}~\sin^2~i$ for $v_{rot}~\sin~i~=~1500~\rm{km~s^{-1}}$. In the literature the inclination is given as $45^{\circ}$ with an error around $10^{\circ}$ (e.g., \citealt{Das2005,Mueller2011}) and thus $\sin^2~i$ leads to a factor of 0.5 and hence radii of the BLR around 15 and 40~ld. These radii lie well within the range in which \cite{Shapovalova2008} found emitting gas (1 to 50~ld) between 1996 and 2006.

According to \cite{Baskin2018} the outer radius of the BLR is located at $1.6~r_{BLR}$. This leads to an outer radius of the BLR or inner dust radii of $66~\pm~15~\rm{ld}$ and $24~\pm~5~\rm{ld}$. Both of these values are slightly lower than the largest and smallest dust radius  found by \cite{Koshida2009} (71 and 33~ld). However, \cite{Schnuelle2013} showed that the determined dust radius depends on the infrared wavelength used for the dust reverberation mapping. As \cite{Koshida2009} used the K band they will not have picked up this innermost dust radius. Therefore our determined $r_{BLR}$ is not in contradiction with their dust radius. The flux present in the peak is responsible for about 5~\% of the total BEL flux and the area in which this flux is produced (with a width of $\sigma_{\varphi,2}~\approx~0.45~\rm{rad}$ and at a smaller radius) is also in the range of 5~\% the area in which the overall BEL was produced. This shows that the area should be large enough to produce the additional flux.

With this radius we can also determine a temperature inside the accretion disk following the line of argument of \cite{Czerny2011} using the monochromatic luminosity at 5100~$\AA{}$. 
We can take this luminosity from \cite{Shapovalova2008} and get a disk temperature of around 450~K at $r_{in}$. However looking at the optical lightcurve in both \cite{Shapovalova2008} and \cite{Koshida2009} of NGC~4151, the optical flux was significantly higher in December~2005 (with no data until June~2005). As the clouds are produced before becoming visible as part of the BEL we can get to temperatures of approximately 600~K easily considering an increased monochromatic luminosity (or potentially even higher if the optical flux would have been even higher between June and December~2005). This temperature is consistent with the dispersion of temperatures found by \cite{Czerny2011}, who published a similar temperature of 550~K for NGC~4151. To summarize, while at the centro-nuclear radii, where we locate the clouds responsible for the BLR shape variation, the temperatures above the accretion disk appear too high for in-situ dust formation,  they drop to values allowing to form dust inside the accretion disk of NGC~4151 at the same radii.

\section{Summary and conclusions}
\label{ch6}

We connected BEL shape variability to the decrease of the dust radius in NGC~4151 between May 2004 and January
2006 in this paper. The results we obtain are:

\begin{enumerate}
 \item The simultaneous decrease of the dust radius and BEL shape variability point to a connection between the BLR and the dust torus. Additionally the velocity, at which the shape variability occurs, indicates a similar decrease of dust radius and BLR radius.
 \item The dust needed for the reduced dust radius is presumably produced in the accretion disk as the temperatures above the accretion disk hardly reach temperatures low enough for dust production. Inflows are also unlikely as a reason for the change of dust radius due to the short timescales of the changes of dust radius. This leaves us with dust production in the accretion disk described in dust inflated accretion disk models. The indications for a similar decrease in radius of both the clouds of the dust torus and BLR provide evidence that the dust and BLR clouds share a similar origin.
 \item The correlated changes in the BEL discussed occur in only a small range of velocities (visible as transient bump). This indicates missing broadening via velocity dispersion of the fresh clouds which can be naturally explained by the here favored formation scenario in an accretion disk wind. We cannot significantly constrain the azimuthal extension of the cloud formation zone but can rule out a completely symmetrical distribution all around the nucleus. If the dust torus and BLR are indeed similar in their production mechanism it is possible that the dust torus shows a similar distribution of clouds.
 \item The location of peaks in BELs can give us information of the radial position of the BLR. In particular if the peak is well defined and sharp the probability is high that the additional clouds are located close to $\varphi~=~\pm~0.5~\pi$. For azimuthal angles with lower projected velocities the peak would be broadened and a much smaller extend in azimuthal direction of the additional clouds would be needed. This explains why similar cloud formation events are less observable if occurring at different azimuthal angles.
\end{enumerate}

As this is only one occasion where we were able to observe the described scenario, we cannot (and do not want to) rule out other dust production outside of the accretion disk in other objects, at other times or at higher radii than those investigated here. For the same reason we cannot speculate what might cause the changes in dust production or if these are just statistical variations. In contrast, our analysis of combined datasets shows evidence supporting dust production in the accretion disk and a similar production mechanism of the dust torus and BLR. For future campaigns, it is important to observe the BLR and dust torus at the same time and with sufficient temporal sampling (bi-weekly for Seyfert-type AGN) in order to robustly detect similar events. Modern extremely-wide-band spectrographs, delivering optical-near-infrared in one shot (like the VLT x-Shooter) are most adequate to achieve this. Such data will help to improve our understanding of dust creation in AGN and how the radius of the dust torus and the BLR changes.

\begin{acknowledgements}

\end{acknowledgements}

\bibliographystyle{aa} 
\bibliography{references} 

\begin{thebibliography}{43}
\expandafter\ifx\csname natexlab\endcsname\relax\def\natexlab#1{#1}\fi

\bibitem[{{Antonucci}(1993)}]{Antonucci1993}
{Antonucci}, R. 1993, \araa, 31, 473

\bibitem[{{Bannikova} {et~al.}(2012){Bannikova}, {Vakulik}, \&
  {Sergeev}}]{Bannikova2012}
{Bannikova}, E.~Y., {Vakulik}, V.~G., \& {Sergeev}, A.~V. 2012, \mnras, 424,
  820

\bibitem[{{Barvainis}(1987)}]{Barvainis1987}
{Barvainis}, R. 1987, \apj, 320, 537

\bibitem[{{Baskin} \& {Laor}(2018)}]{Baskin2018}
{Baskin}, A. \& {Laor}, A. 2018, \mnras, 474, 1970

\bibitem[{{Bentz} {et~al.}(2006){Bentz}, {Peterson}, {Pogge}, {Vestergaard}, \&
  {Onken}}]{Bentz2006}
{Bentz}, M.~C., {Peterson}, B.~M., {Pogge}, R.~W., {Vestergaard}, M., \&
  {Onken}, C.~A. 2006, \apj, 644, 133

\bibitem[{{Chartas} {et~al.}(2003){Chartas}, {Brandt}, \&
  {Gallagher}}]{Chartas2003}
{Chartas}, G., {Brandt}, W.~N., \& {Gallagher}, S.~C. 2003, \apj, 595, 85

\bibitem[{{Cushing} {et~al.}(2004){Cushing}, {Vacca}, \&
  {Rayner}}]{Cushing2004}
{Cushing}, M.~C., {Vacca}, W.~D., \& {Rayner}, J.~T. 2004, \pasp, 116, 362

\bibitem[{{Czerny} \& {Hryniewicz}(2011)}]{Czerny2011}
{Czerny}, B. \& {Hryniewicz}, K. 2011, \aap, 525, L8

\bibitem[{{Czerny} {et~al.}(2017){Czerny}, {Li}, {Hryniewicz}, {Panda},
  {Wildy}, {Sniegowska}, {Wang}, {Sredzinska}, \& {Karas}}]{Czerny2017}
{Czerny}, B., {Li}, Y.-R., {Hryniewicz}, K., {et~al.} 2017, \apj, 846, 154

\bibitem[{{Das} {et~al.}(2005){Das}, {Crenshaw}, {Hutchings}, {Deo}, {Kraemer},
  {Gull}, {Kaiser}, {Nelson}, \& {Weistrop}}]{Das2005}
{Das}, V., {Crenshaw}, D.~M., {Hutchings}, J.~B., {et~al.} 2005, \aj, 130, 945

\bibitem[{{Eracleous} \& {Halpern}(2003)}]{Eracleous2003}
{Eracleous}, M. \& {Halpern}, J.~P. 2003, \apj, 599, 886

\bibitem[{{Grier} {et~al.}(2013){Grier}, {Martini}, {Watson}, {Peterson},
  {Bentz}, {Dasyra}, {Dietrich}, {Ferrarese}, {Pogge}, \& {Zu}}]{Grier2013}
{Grier}, C.~J., {Martini}, P., {Watson}, L.~C., {et~al.} 2013, \apj, 773, 90

\bibitem[{{Groenewegen} {et~al.}(2009){Groenewegen}, {Sloan}, {Soszy{\'n}ski},
  \& {Petersen}}]{Groenewegen2009}
{Groenewegen}, M.~A.~T., {Sloan}, G.~C., {Soszy{\'n}ski}, I., \& {Petersen},
  E.~A. 2009, \aap, 506, 1277

\bibitem[{{H{\"o}nig} \& {Kishimoto}(2010)}]{Hoenig2010}
{H{\"o}nig}, S.~F. \& {Kishimoto}, M. 2010, \aap, 523, A27

\bibitem[{{H{\"o}nig} {et~al.}(2017){H{\"o}nig}, {Watson}, {Kishimoto},
  {Gandhi}, {Goad}, {Horne}, {Shankar}, {Banerji}, {Boulderstone}, {Jarvis},
  {Smith}, \& {Sullivan}}]{Hoenig2017}
{H{\"o}nig}, S.~F., {Watson}, D., {Kishimoto}, M., {et~al.} 2017, \mnras, 464,
  1693

\bibitem[{{Ili{\'c}} {et~al.}(2015){Ili{\'c}}, {Popovi{\'c}}, {Shapovalova},
  {Burenkov}, {Chavushyan}, \& {Kova{\v c}evi{\'c}}}]{Ilic2015}
{Ili{\'c}}, D., {Popovi{\'c}}, L.~{\v C}., {Shapovalova}, A.~I., {et~al.} 2015,
  Journal of Astrophysics and Astronomy, 36, 433

\bibitem[{{Kaspi} {et~al.}(2000){Kaspi}, {Smith}, {Netzer}, {Maoz}, {Jannuzi},
  \& {Giveon}}]{Kaspi2000}
{Kaspi}, S., {Smith}, P.~S., {Netzer}, H., {et~al.} 2000, \apj, 533, 631

\bibitem[{{Kishimoto} {et~al.}(2007){Kishimoto}, {H{\"o}nig}, {Beckert}, \&
  {Weigelt}}]{Kishimoto2007}
{Kishimoto}, M., {H{\"o}nig}, S.~F., {Beckert}, T., \& {Weigelt}, G. 2007,
  \aap, 476, 713

\bibitem[{{Koshida} {et~al.}(2009){Koshida}, {Yoshii}, {Kobayashi}, {Minezaki},
  {Sakata}, {Sugawara}, {Enya}, {Suganuma}, {Tomita}, {Aoki}, \&
  {Peterson}}]{Koshida2009}
{Koshida}, S., {Yoshii}, Y., {Kobayashi}, Y., {et~al.} 2009, \apjl, 700, L109

\bibitem[{{Landt} {et~al.}(2011{\natexlab{a}}){Landt}, {Bentz}, {Peterson},
  {Elvis}, {Ward}, {Korista}, \& {Karovska}}]{Landt2011b}
{Landt}, H., {Bentz}, M.~C., {Peterson}, B.~M., {et~al.} 2011{\natexlab{a}},
  \mnras, 413, L106

\bibitem[{{Landt} {et~al.}(2008){Landt}, {Bentz}, {Ward}, {Elvis}, {Peterson},
  {Korista}, \& {Karovska}}]{Landt2008}
{Landt}, H., {Bentz}, M.~C., {Ward}, M.~J., {et~al.} 2008, \apjs, 174, 282

\bibitem[{{Landt} {et~al.}(2011{\natexlab{b}}){Landt}, {Elvis}, {Ward},
  {Bentz}, {Korista}, \& {Karovska}}]{Landt2011}
{Landt}, H., {Elvis}, M., {Ward}, M.~J., {et~al.} 2011{\natexlab{b}}, \mnras,
  414, 218

\bibitem[{{Landt} {et~al.}(2015){Landt}, {Ward}, {Steenbrugge}, \&
  {Ferland}}]{Landt2015}
{Landt}, H., {Ward}, M.~J., {Steenbrugge}, K.~C., \& {Ferland}, G.~J. 2015,
  \mnras, 449, 3795

\bibitem[{{Lira} {et~al.}(2018){Lira}, {Botti}, {Kaspi}, \&
  {Netzer}}]{Lira2018}
{Lira}, P., {Botti}, I., {Kaspi}, S., \& {Netzer}, H. 2018, ArXiv e-prints

\bibitem[{{M{\"u}ller-S{\'a}nchez} {et~al.}(2011){M{\"u}ller-S{\'a}nchez},
  {Prieto}, {Hicks}, {Vives-Arias}, {Davies}, {Malkan}, {Tacconi}, \&
  {Genzel}}]{Mueller2011}
{M{\"u}ller-S{\'a}nchez}, F., {Prieto}, M.~A., {Hicks}, E.~K.~S., {et~al.}
  2011, \apj, 739, 69

\bibitem[{{Nenkova} {et~al.}(2002){Nenkova}, {Ivezi{\'c}}, \&
  {Elitzur}}]{Nenkova2002}
{Nenkova}, M., {Ivezi{\'c}}, {\v Z}., \& {Elitzur}, M. 2002, \apjl, 570, L9

\bibitem[{{Nenkova} {et~al.}(2008){Nenkova}, {Sirocky}, {Ivezi{\'c}}, \&
  {Elitzur}}]{Nenkova2008}
{Nenkova}, M., {Sirocky}, M.~M., {Ivezi{\'c}}, {\v Z}., \& {Elitzur}, M. 2008,
  \apj, 685, 147

\bibitem[{{Rayner} {et~al.}(2003){Rayner}, {Toomey}, {Onaka}, {Denault},
  {Stahlberger}, {Vacca}, {Cushing}, \& {Wang}}]{Rayner2003}
{Rayner}, J.~T., {Toomey}, D.~W., {Onaka}, P.~M., {et~al.} 2003, \pasp, 115,
  362

\bibitem[{{Riffel} {et~al.}(2006){Riffel}, {Rodr{\'{\i}}guez-Ardila}, \&
  {Pastoriza}}]{Riffel2006}
{Riffel}, R., {Rodr{\'{\i}}guez-Ardila}, A., \& {Pastoriza}, M.~G. 2006, \aap,
  457, 61

\bibitem[{{Schartmann} {et~al.}(2010){Schartmann}, {Burkert}, {Krause},
  {Camenzind}, {Meisenheimer}, \& {Davies}}]{Schartmann2010}
{Schartmann}, M., {Burkert}, A., {Krause}, M., {et~al.} 2010, \mnras, 403, 1801

\bibitem[{{Schartmann} {et~al.}(2005){Schartmann}, {Meisenheimer}, {Camenzind},
  {Wolf}, \& {Henning}}]{Schartmann2005}
{Schartmann}, M., {Meisenheimer}, K., {Camenzind}, M., {Wolf}, S., \&
  {Henning}, T. 2005, \aap, 437, 861

\bibitem[{{Schn{\"u}lle} {et~al.}(2013){Schn{\"u}lle}, {Pott}, {Rix},
  {Decarli}, {Peterson}, \& {Vacca}}]{Schnuelle2013}
{Schn{\"u}lle}, K., {Pott}, J.-U., {Rix}, H.-W., {et~al.} 2013, \aap, 557, L13

\bibitem[{{Schn{\"u}lle} {et~al.}(2015){Schn{\"u}lle}, {Pott}, {Rix},
  {Peterson}, {De Rosa}, \& {Shappee}}]{Schnuelle2015}
{Schn{\"u}lle}, K., {Pott}, J.-U., {Rix}, H.-W., {et~al.} 2015, \aap, 578, A57

\bibitem[{{Shapovalova} {et~al.}(2010){Shapovalova}, {Popovi{\'c}}, {Burenkov},
  {Chavushyan}, {Ili{\'c}}, {Kova{\v c}evi{\'c}}, {Bochkarev}, \&
  {Le{\'o}n-Tavares}}]{Shapovalova2010}
{Shapovalova}, A.~I., {Popovi{\'c}}, L.~{\v C}., {Burenkov}, A.~N., {et~al.}
  2010, \aap, 509, A106

\bibitem[{{Shapovalova} {et~al.}(2008){Shapovalova}, {Popovi{\'c}}, {Collin},
  {Burenkov}, {Chavushyan}, {Bochkarev}, {Ben{\'{\i}}tez}, {Dultzin}, {Kova{\v
  c}evi{\'c}}, {Borisov}, {Carrasco}, {Le{\'o}n-Tavares}, {Mercado}, {Valdes},
  {Vlasuyk}, \& {Zhdanova}}]{Shapovalova2008}
{Shapovalova}, A.~I., {Popovi{\'c}}, L.~{\v C}., {Collin}, S., {et~al.} 2008,
  \aap, 486, 99

\bibitem[{{Stern} {et~al.}(2015){Stern}, {Hennawi}, \& {Pott}}]{Stern2015}
{Stern}, J., {Hennawi}, J.~F., \& {Pott}, J.-U. 2015, \apj, 804, 57

\bibitem[{{Suganuma} {et~al.}(2006){Suganuma}, {Yoshii}, {Kobayashi},
  {Minezaki}, {Enya}, {Tomita}, {Aoki}, {Koshida}, \&
  {Peterson}}]{Suganuma2006}
{Suganuma}, M., {Yoshii}, Y., {Kobayashi}, Y., {et~al.} 2006, \apj, 639, 46

\bibitem[{{Sulentic} {et~al.}(2000){Sulentic}, {Marziani}, \&
  {Dultzin-Hacyan}}]{Sulentic2000}
{Sulentic}, J.~W., {Marziani}, P., \& {Dultzin-Hacyan}, D. 2000, \araa, 38, 521

\bibitem[{{Tombesi} {et~al.}(2010){Tombesi}, {Cappi}, {Reeves}, {Palumbo},
  {Yaqoob}, {Braito}, \& {Dadina}}]{Tombesi2010}
{Tombesi}, F., {Cappi}, M., {Reeves}, J.~N., {et~al.} 2010, \aap, 521, A57

\bibitem[{{Urry} \& {Padovani}(1995)}]{Urry1995}
{Urry}, C.~M. \& {Padovani}, P. 1995, \pasp, 107, 803

\bibitem[{{Vacca} {et~al.}(2003){Vacca}, {Cushing}, \& {Rayner}}]{Vacca2003}
{Vacca}, W.~D., {Cushing}, M.~C., \& {Rayner}, J.~T. 2003, \pasp, 115, 389

\bibitem[{{Watson} {et~al.}(2011){Watson}, {Denney}, {Vestergaard}, \&
  {Davis}}]{Watson2011}
{Watson}, D., {Denney}, K.~D., {Vestergaard}, M., \& {Davis}, T.~M. 2011,
  \apjl, 740, L49

\bibitem[{{Wildy} {et~al.}(2016){Wildy}, {Landt}, {Goad}, {Ward}, \&
  {Collinson}}]{Wildy2016}
{Wildy}, C., {Landt}, H., {Goad}, M.~R., {Ward}, M., \& {Collinson}, J.~S.
  2016, \mnras, 461, 2085

\end{thebibliography}

\end{document}